\documentclass[9pt,twocolumn,twoside]{opticajnl}
\journal{opticajournal} % use for journal or Optica Open submissions
\usepackage[utf8]{inputenc}
\DeclareUnicodeCharacter{03BB}{$\lambda$}
\DeclareUnicodeCharacter{03C0}{$\pi$}
\setboolean{shortarticle}{false}
\usepackage{subcaption}
\makeatletter
\renewcommand*\subcaption@label{\caption@withoptargs\subcaption@@label}
\makeatother
\usepackage{stfloats}
\usepackage{multirow}
\usepackage{siunitx}
\usepackage{physics}
\DeclareSIUnit\baud{Baud}

\usepackage{lineno}
\title{Chip-Based 16 GBaud Continuous-Variable Quantum Key Distribution}

\author[1,$\dagger$, *]{Adnan A.E. Hajomer}
\author[2,$\dagger$]{Axl Bomhals}
\author[2]{C\'{e}dric Bruynsteen}
\author[2]{Aboobackkar Sidhique}
\author[3]{Ivan Derkach}
\author[1]{Ulrik L. Andersen}
\author[2, *]{Xin Yin}
\author[1,*]{Tobias Gehring}

\affil[1]{Center for Macroscopic Quantum States (bigQ), Department of Physics, Technical University of Denmark, 2800 Kongens Lyngby, Denmark}
\affil[2]{Ghent University-imec, IDLab, Dep. INTEC, 9052 Ghent, Belgium}
\affil[3]{Department of Optics, Faculty of Science, Palacky University, 17. listopadu 12, 771 46 Olomouc, Czech Republic}
\affil[$\dagger$]{These authors contributed equally}
\affil[*]{Corresponding authors: aaeha@dtu.dk, tobias.gehring@fysik.dtu.dk, xin.yin@ugent.be}

\begin{abstract}
Quantum key distribution (QKD) stands as the most successful application of quantum information science, providing information-theoretic security for key exchange. While it has evolved from proof-of-concept experiments to commercial products, widespread adoption requires chip-based integration to reduce costs, enable mass production, facilitate miniaturization, and enhance system performance. Here, we demonstrate the first fully photonic-integrated continuous-variable QKD (CVQKD) system operating at a classical telecom symbol rate of 16 GBaud. Our system integrates a silicon photonic transmitter circuit (excluding the laser source) and a 20 GHz photonic-electronic receiver, which features a phase-diverse silicon photonic integrated circuit and custom-designed GaAs pHEMT transimpedance amplifiers. Advanced digital signal processing allows our system to achieve the highest reported secure key rate to date, reaching 0.289 Gb/s and 0.246 Gb/s over a 20 km fiber link in the asymptotic and finite-size regimes, respectively. These results establish a record key rate and represent a critical step toward scalable, cost-effective, and mass-deployable quantum-secure communication using photonic-integrated CVQKD systems. 
\end{abstract}

\setboolean{displaycopyright}{false} % Do not include copyright or licensing information in submission.

\begin{document}

\maketitle

\section{Introduction}
Quantum key distribution (QKD), the cornerstone of quantum communication, enables the secure exchange of cryptographic keys over public channels with information-theoretic security~\cite{BEN84, ekert1991quantum}, making it resistant to attacks from both classical and quantum computers. While QKD has transitioned from a theoretical concept to real-world implementations, including field trials and commercial products~\cite{pirandola2020advances}, its widespread adoption is still hindered by scalability, network integration challenges, and high manufacturing costs~\cite{diamanti2016practical}. Photonic integration is a key enabler for overcoming these limitations, allowing for miniaturization, cost reduction, mass production, and enhanced system performance~\cite{luo2023recent}—particularly in achieving high secure key rates necessary for modern data-intensive applications.

Among various QKD implementations, continuous-variable QKD (CVQKD) is particularly well-suited for photonic integration due to its inherent compatibility with standard telecommunications technologies~\cite{usenko2025continuous}. In CVQKD, information is encoded in the continuous degrees of freedom of a quantum system-such as the amplitude and phase quadratures of the electromagnetic field of light— using a quadrature modulator and decoded via coherent detection~\cite{grosshans2002continuous}. These core operations of encoding and decoding can be efficiently implemented in integrated photonic circuits fabricated using standard foundry processes. Furthermore, CVQKD can achieve high secure key rates, approaching the Pirandola-Laurenza-Ottaviani-Banchi (PLOB) bound~\cite{pirandolaFundamentalLimitsRepeaterless2017}, making it well-suited for high-data-rate quantum-secured communication. However, despite its intrinsic compatibility with photonic integration, most prior efforts have focused on integrating individual system components rather than developing a fully integrated CVQKD solution~\cite{li2023continuous,bian2024continuous,bian2024highly,pietri2024experimental,Aldama:25}. While Zhang et al.~\cite{zhang2019integrated} demonstrated a chip-based CVQKD system without an integrated laser source, their secure key rate limited to 250 kbit/s due to the limited system bandwidth. This underscores the pressing need for high-performance, fully integrated CVQKD systems that can achieve higher secure key rates for practical deployment.
\par In this work, we report the fastest fully integrated CVQKD system to date (excluding the laser source), operating at a symbol rate of 16 GBaud. Our photonic-integrated transmitter incorporates a silicon photonic in-phase and quadrature (IQ) modulator with  6~dB bandwidth of 13~GHz, enabling high-speed quantum state encoding. Meanwhile, our photonic-electronic-integrated receiver features a silicon photonic phase-diverse coherent detector, complemented by custom-designed GaAs pHEMT transimpedance amplifiers, achieving quantum noise-limited detection across a 20 GHz bandwidth. By combining these integrated subsystems with advanced digital signal processing (DSP) techniques, including pre- and post-equalization~\cite{hajomer2024continuous}, we achieve record-breaking secure key rates of 0.289 Gb/s in the asymptotic regime and 0.246 Gb/s in the finite-size regime over a 20 km fiber link.
These results mark a significant step forward in the development of scalable, high-speed, and cost-effective CVQKD systems, demonstrating the transformative potential of photonic integration in enabling the mass deployment of quantum-secured communication networks.

\begin{figure}[t]
   \centering
        \includegraphics[width=1\linewidth]{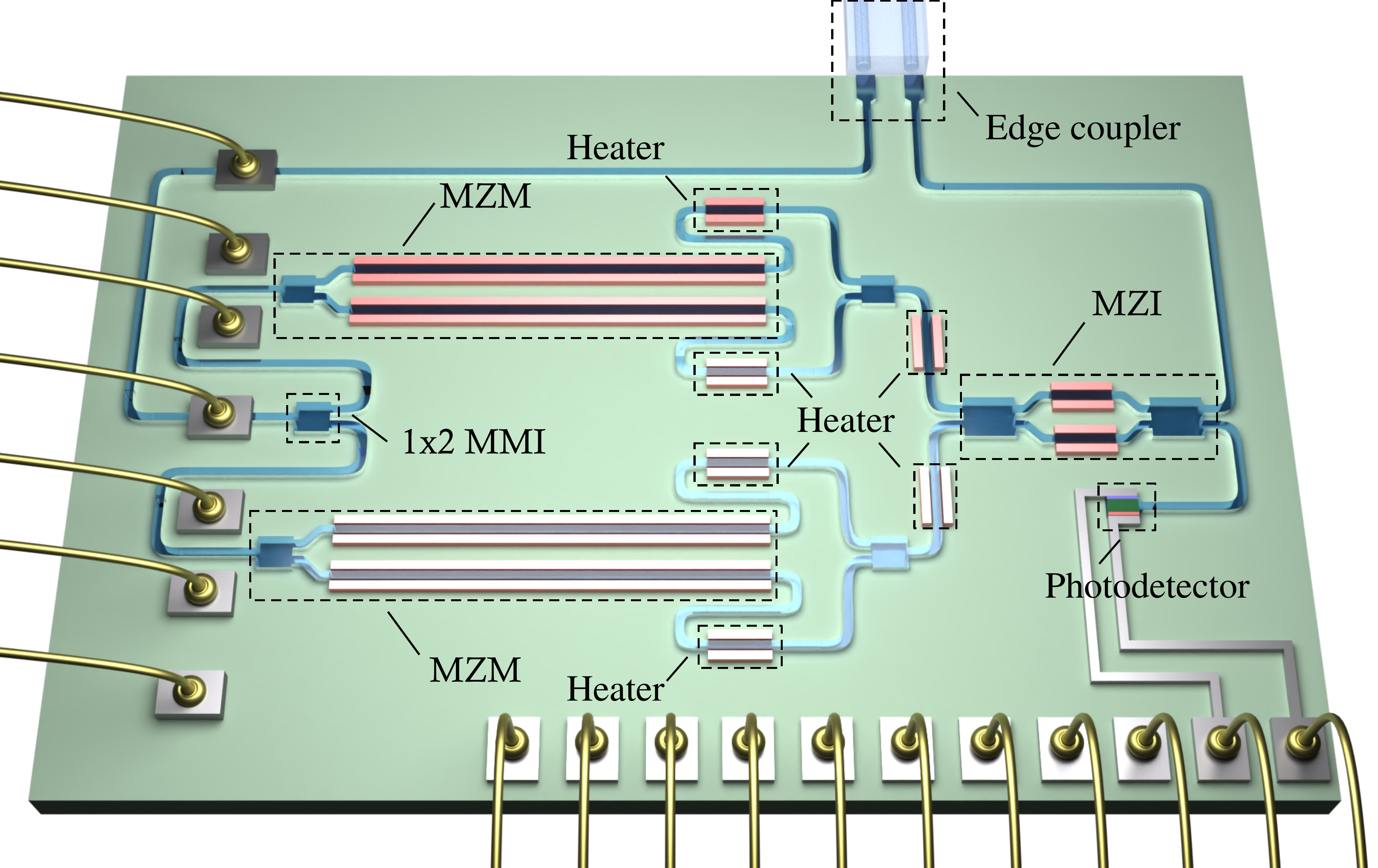}
    \caption{Functional diagram of the IQ modulator photonic integrated circuit. MZM: Mach-Zehnder modulator; MZI: Mach-Zehnder interferometer; MMI: multimode interferometer.}
    \label{fig:TxPic}
\end{figure}

\begin{figure}[t]
   \centering
        \includegraphics[width=1\linewidth]{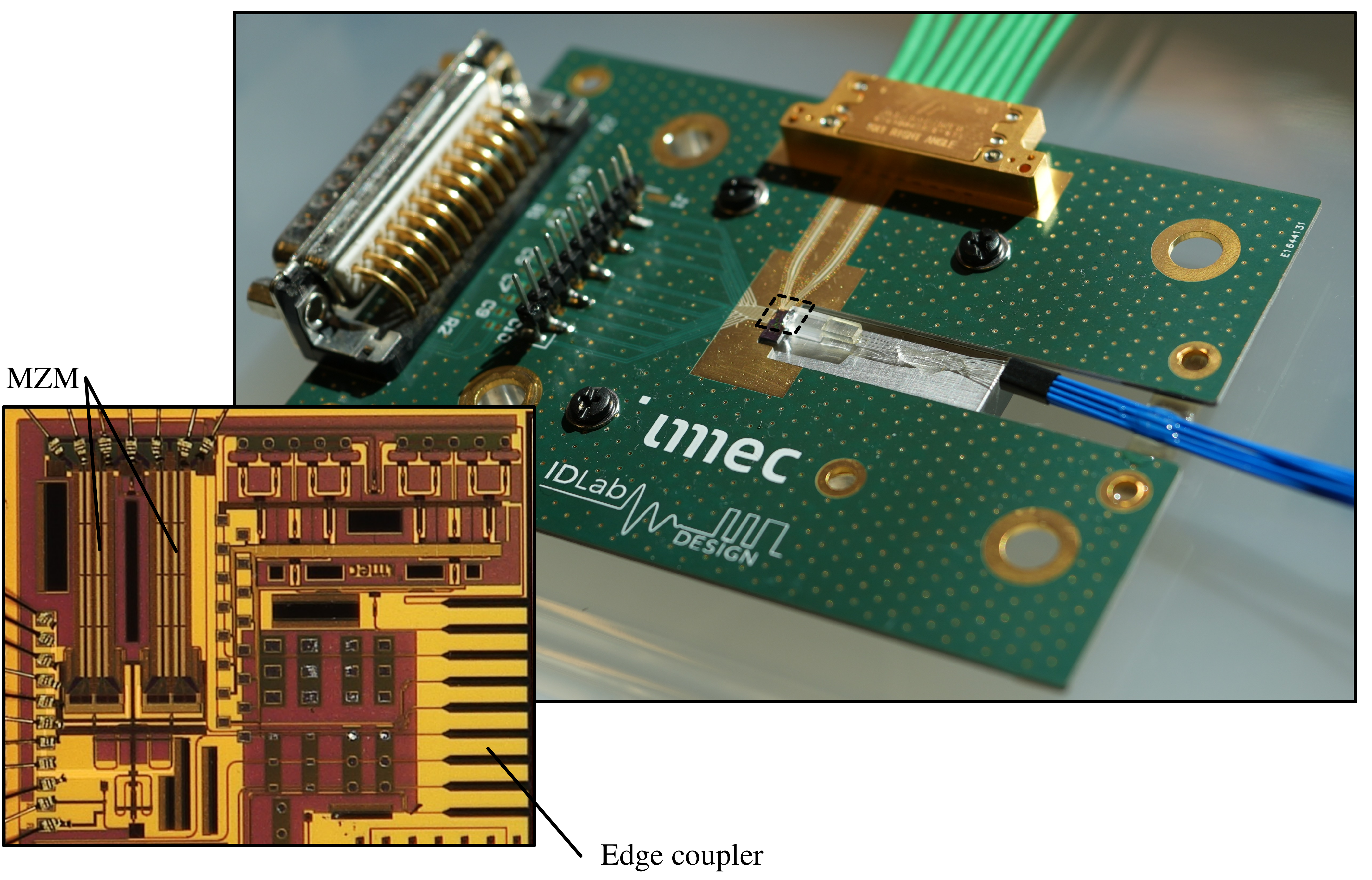}
    \caption{Micrograph of the IQ modulator photonic integrated circuit and its printed circuit board. MZM: Mach-Zehnder modulator.}
    \label{fig:IQAssembly}
\end{figure}

\section{Integrated CVQKD system}

Our integrated CVQKD system comprises an integrated transmitter and receiver fabricated using imec’s ISiPP50G silicon platform. Figure~\ref{fig:TxPic} illustrates the photonic integrated circuit (PIC) of our transmitter, highlighting its core components: an IQ modulator for optical signal modulation, a Mach-Zehnder interferometer (MZI) designed for adjusting IQ amplitude imbalance, and a photodiode for signal monitoring.

\begin{figure}[t]
   \centering
        \includegraphics[width=1\linewidth]{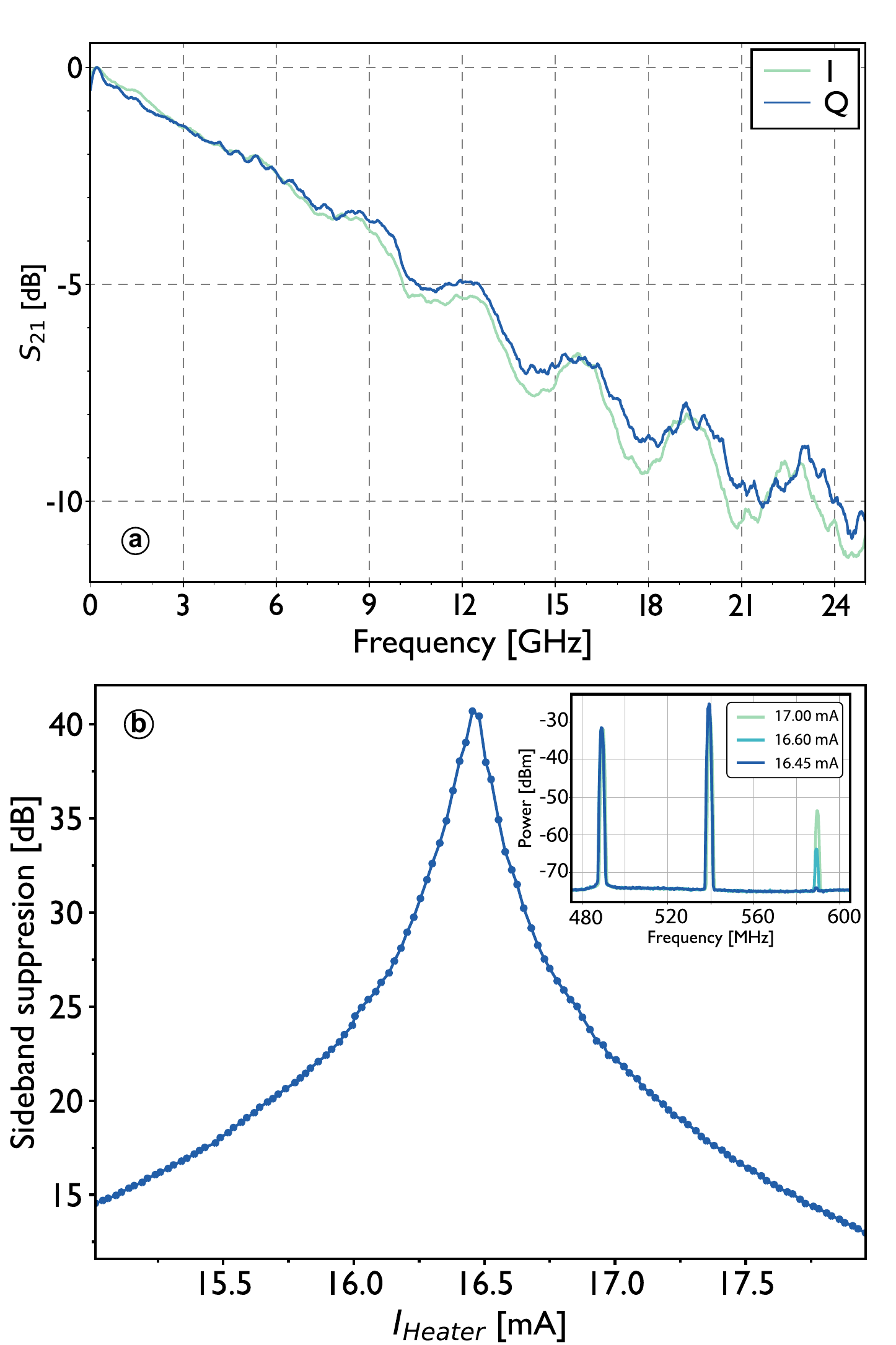}
    \caption{Characterization of the IQ modulator: \textbf{(a)} Frequency response of the nested MZMs obtained from the $\text{S}_{21}$ parameter. \textbf{(b)} Sideband suppression versus heater driving current. The inset displays the corresponding spectra for three representative suppression levels: 40dB, 22dB, and 28~dB.}
    \label{fig:spec}
\end{figure}

The IQ modulator consists of two Mach-Zehnder modulators (MZMs) with a $V_{\pi}$ of 1.1 V nested within a larger MZI structure, utilizing multi-mode interferometers (MMIs) for efficient light splitting and combining. Each MZM incorporates a depletion-mode phase shifter optimized for high-frequency operation, along with dual heaters for precise bias point control, operating together in a push-pull configuration. Furthermore, to achieve the required 90-degree phase shift between the I and Q branches, additional heaters are integrated within the global MZI structure. Fig.~\ref{fig:spec} (a) shows the $\text{S}_{21}$ measurement of the nested MZMs, revealing a similar frequency response and a 6dB bandwidth of 13~GHz.

 Our transmitter can operate in various modes depending on the driving bias voltages and RF signals. A particularly important mode for CVQKD is the optical single-sideband suppressed-carrier mode~\cite{hajomer2024long,jain2022practical}. Ideally, when both MZMs are biased at the dark fringe, maintaining a 90-degree phase shift and equal amplitude for both I and Q quadratures, the carrier and one of the sidebands can be effectively suppressed. However, maintaining a perfect amplitude balance between the I and Q branches is challenging due to manufacturing tolerances. This imbalance can degrade sideband suppression performance, potentially resulting in a security loophole~\cite{jain2021modulation,hajomer2022modulation}. Additionally, it distorts the symmetry of the constellation in phase space, complicating phase carrier recovery and degrading system performance~\cite{hajomer2025finite}. To address this issue, we integrated an additional MZI with a pair of heaters specifically designed to correct amplitude imbalances between the two modulated quadratures. As shown in Fig.~\ref{fig:spec} (b), by tuning these heaters, a sideband suppression of 40.7 dB @ 500 MHz was achieved. Additionally, a photodiode is included to monitor and stabilize the MZI operation. 

A printed circuit board, shown in Fig.~\ref{fig:IQAssembly}, is used to make electrical connections to the PIC for RF and DC signals. These connections are wire-bonded to the PIC. The DC bias voltages for two MZM heaters and a 90-degree phase-shifting heater are connected to a DB25 connector for automatic bias control. The DC traces from the fourth heater in the MZI are connected to headers for manual tuning, along with the anode and cathode traces of the monitor photo detector. The RF inputs to the modulator are wire-bonded to matched differential traces on the PCB and utilize a TR40 connector for external connection. To couple light into and out of the edge couplers of the PIC transmitter, a polarization-maintained fiber array is employed. Two additional edge couplers, connected to a known optical channel, are used to align the fiber array with the help of a six-axis stage. The fiber array was glued with a low-loss adhesive for mechanical stability. The insertion loss of the edge coupler is 8 dB, which is still quite high. With improvements in the alligment process, especially during the curing of the glue, this loss could be further reduced to 3.5 dB seen before curing. However, these losses are not critical to the system, as the output of the modulator is strongly attenuated to achieve the desired modulation variance.    

Our phase-diverse optical front-end, also fabricated using imec’s iSiPP50G silicon photonics platform, couples light into the chip via two grating couplers, splitting the quantum signal and LO into two arms using a 1x2 MMI to measure conjugate quadratures. Two waveguide heaters ensure a 90° phase shift between quadratures, followed by balanced homodyne detectors composed of a MZI and high-responsivity photodetectors. Each pair of balanced photodetectors was connected to a separate TIA, custom-designed in a 100nm GaAs pHEMT process. The integrated photonic and electronic receiver provides a shot-noise limited bandwidth of 20 GHz and achieves an efficiency of 44\%. For further details about the optical front-end design and integrated transimpedance amplifier of our receiver, we refer the reader to Ref.~\cite{hajomer2024continuous,Bruynsteen:21}.

%\textcolor{red}{ Axl, Cedric and Abo}

\begin{figure*}[t]
   \centering
    \includegraphics[width=1\linewidth]{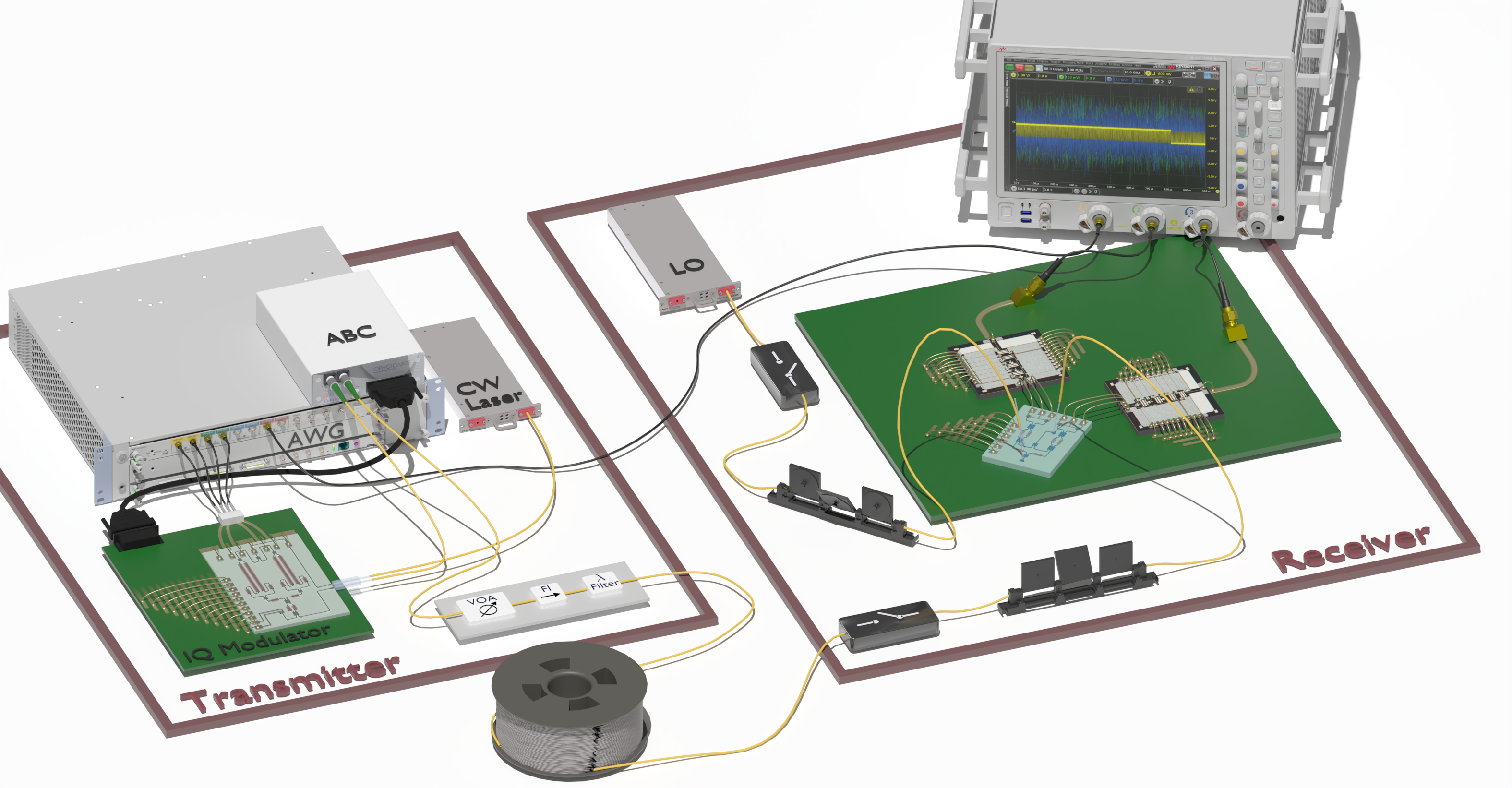}
    \caption{CVQKD system layout highlighting essential transmitter and receiver components. At the transmitter side: CW laser (continuous wave laser), IQ modulator (in-phase and quadrature modulator) with edge coupler on an interposer PCB (dark green box), ABC (automatic bias controller), VOA (variable optical attenuator), FI (Faraday isolator), wavelength filter, and AWG (arbitrary waveform generator). At the receiver side: CW laser used as a local oscillator (LO), two optical switches, two polarization controllers, the photonic and electronic integrated circuits on an interposer PCB (dark green box), and RTO (real-time oscilloscope).}
    \label{fig:SetupPic}
\end{figure*}

\section{High-rate CVQKD SYSTEM}

Figure~\ref{fig:SetupPic} presents our high-speed, chip-based CVQKD system, which implements a discrete-modulated (DM) CVQKD protocol~\cite{denys2021explicit}. The system consists of a transmitter (Alice) and a receiver (Bob), connected through a quantum channel composed of standard single-mode fiber (SSMF). The following sections provide a detailed description of the system components.

\subsection{Transmitter (Alice)}
At Alice's station, a 1550 nm continuous-wave (CW) laser with a narrow linewidth of 100 Hz served as the optical source. The coherent states were generated using our integrated IQ modulator, which was driven by an 8-bit arbitrary waveform generator (AWG, Keysight M8195A) operating at a sampling rate of 32 GSa/s. The bias voltages of the modulator were controlled by a commercial automatic bias controller (ABC). The modulation variance of the transmitted quantum states was adjusted using a variable optical attenuator (VOA) and by tuning the AWG driving voltage. To prevent back-reflections and mitigate potential Trojan-horse attacks, a Faraday isolator and optical filter were placed at the transmitter’s output.

The complex amplitude of each coherent state, $\ket{\alpha}=\ket{x+jp}$, was randomly selected at a rate of 16 GBaud from a probabilistically shaped discrete constellation with a predefined modulation order (M) and a Gaussian-like probability distribution. The quantum symbols were upsampled to match the AWG sampling rate and pulse-shaped using a root-raised cosine (RRC) filter with a roll-off factor of 0.2. To support the high symbol rates, a pre-emphasis filter was applied to compensate for the high-frequency roll-off of the transmitter, with coefficients derived from the inverse transfer function of the transmitter~\cite{hajomer2024continuous}. Additionally, a 15 GHz pilot tone was frequency-multiplexed with the quantum signal to establish a phase reference between the transmitter and receiver.

\subsection{Receiver (Bob)}
At the receiver, the coherent states were measured using an intradyne detection scheme ~\cite{hajomer2024continuous,kikuchi2015fundamentals}. This scheme employed our integrated phase-diverse receiver along with a free-running CW laser serving as the local oscillator (LO), which operated independently of Alice’s laser. To avoid degradation from vacuum mode in the image sideband, the frequency offset between the two lasers was maintained below half the quantum signal bandwidth. Two polarization controllers (PCs) were used to align the LO and the quantum signal for optimal coupling into the receiver chip. Additionally, two optical switches were incorporated in the LO and signal paths to enable vacuum noise, and electronic noise calibration. Following detection, the signal was digitized using an 8-bit real-time oscilloscope (RTO, Keysight DSA Z634A) operating at 80 GSa/s, synchronized with the AWG via a 100 MHz reference clock. The digitized signals were subsequently processed offline.

To recover the quantum symbols, an initial frequency-domain equalizer (whitening filter) was applied to compensate for the high-frequency roll-off of the receiver. The filter coefficients were determined from the inverse frequency response obtained via vacuum noise measurements. The frequency offset between Alice’s and Bob’s lasers was estimated using the Hilbert transform of the pilot tone, followed by a linear phase fit of the extracted phase profile. The relative phase was then determined by shifting the pilot tone to baseband using the estimated frequency offset. The quantum signal was then transformed to baseband and corrected for phase. A cross-correlation between the transmitted and received reference samples was employed to correct for propagation delays introduced by fiber transmission and electronic components. Finally, after matched filtering and downsampling, the quantum symbols were reconstructed for further processing.

\section{Results}
The security of CVQKD is evaluated using the Devetak-Winters bound, quantifying the information advantage that trusted parties (Alice and Bob) have over an eavesdropper (Eve) controlling the quantum channel. The bound is defined as $R=\beta I(A:B)-\chi(E:B)$, where $I(A:B)$ is the mutual information between Alice and Bob, calculated directly from their encoded and measured data, and closely related to the signal-to-noise ratio (SNR). The mutual information is limited by the efficiency of information reconciliation $\beta\in\left[0,1\right]$. The second term, $\chi(E:B)$, is the Holevo bound, representing the maximum information Eve can obtain considering collective attacks~\cite{usenko2025continuous}. Under the assumption of a Gaussian quantum channel, $\chi(E:B)$ can be evaluated from two experimentally estimated parameters: the channel loss $\eta$ and the excess noise $\varepsilon$.

A key system parameter controlled by the sender is the modulation variance $V_M$, defining both the alphabet size and the resulting SNR. Optimizing $V_M$ is crucial for achieving a positive key rate $R$  under practical reconciliation efficiency $\beta<1$~\cite{hajomer2024long}. However,  increasing $V_M$  also amplifies residual phase noise ~\cite{hajomer2024long,hajomer2024QPON}, limiting how large it can be set. Conversely, a low  $V_M$ reduces the phase noise and—crucially—allows the protocol to closely approximate  the performance of Gaussian-modulated CVQKD using a finite number of discrete states displaced according to a continuous, zero-mean Gaussian distribution~\cite{denys2021explicit}. \par

\begin{figure*}[t]
    \centering
    \includegraphics[width=0.99\linewidth]{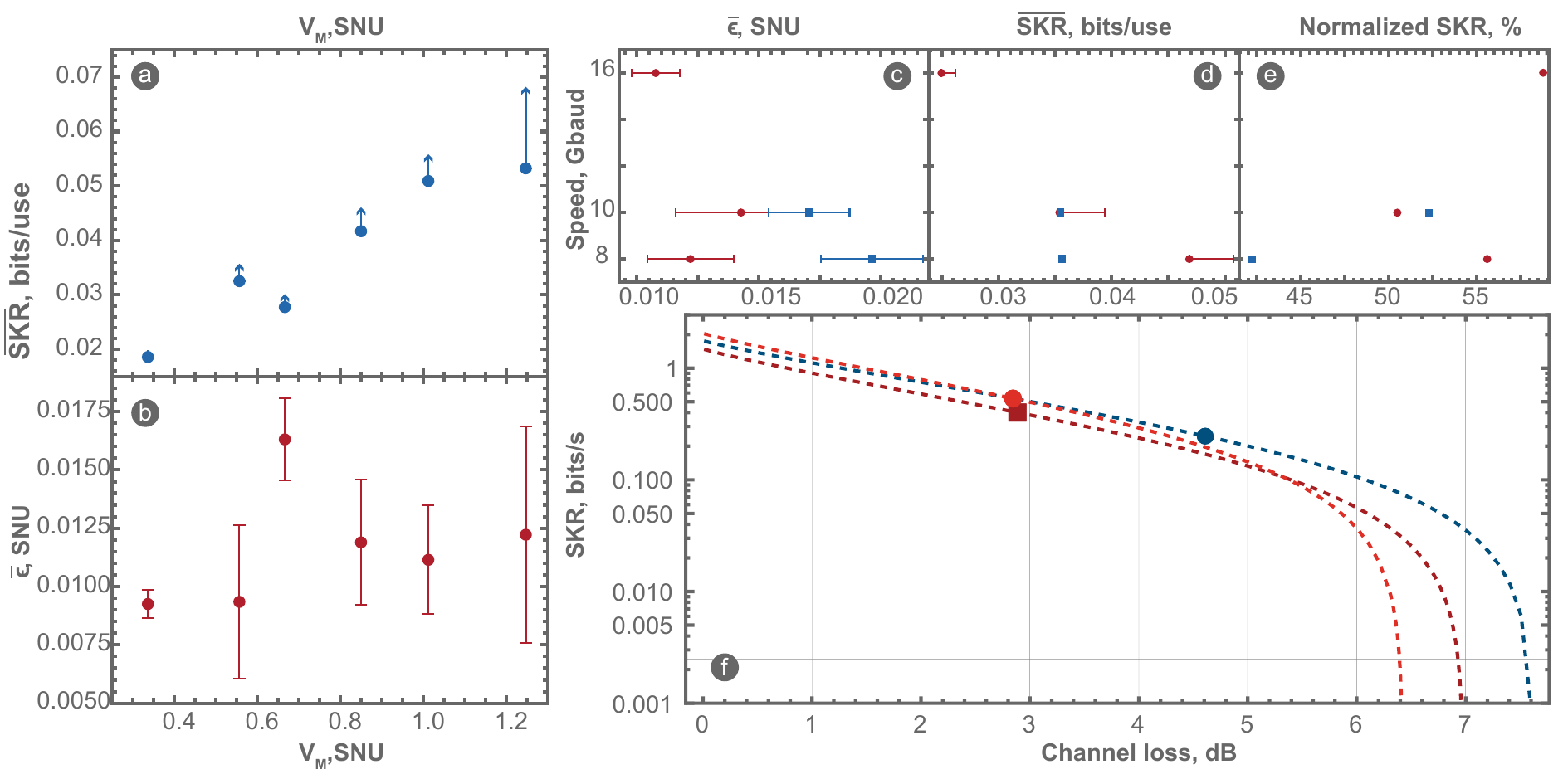}
    \caption{Security analysis results. (\textbf{a}-\textbf{b}) Averaged (over respective abscissa parameter) experimental results for 37 measurements over a 10 km fiber channel. (\textbf{a}) Asymptotic secure key rate in bits per channel use (arrows indicate maximal theoretical performance achieved with continuous Gaussian CV QKD protocol) dependency on the modulation variance $V_M$, and (\textbf{b}) respective averaged excess noise $ \varepsilon$. (\textbf{c-e}) Influence of operation speed on (\textbf{c}) excess noise $ \varepsilon$ and secure key rate with finite-size parameter estimation in (\textbf{d}) bits per channel use, and (\textbf{e}) average key rate at given speed normalized to the highest achieved key generation speed. (\textbf{f}) Three measurement results over 10 and 20 km fibers, with theoretical lines indicating theoretical projected performance with the same excess noise at the channel input. Circles show constellation of size $M=16$ and $\nu_{16}=0.215$, squares $M=64$ and $\nu_{64}=0.129$. Reconciliation efficiency $\beta=95\%$.}
    \label{fig:results-ALL}
\end{figure*}

In this work, we operate our system in the low modulation variance regime, \textit{i.e.}, $V_M \lesssim 1$ shot-noise unit (SNU), using discrete, probabilistically-shaped constellations of 16 and 64 states. These constellations are respectively optimized for modulation variances around $V_M \approx 1$ SNU. Security is first evaluated in the asymptotic regime, using an analytical bound for arbitrary modulations under the Gaussian channel assumption~\cite{denys2021explicit}.

We conduct 37 measurements over a 10~km fiber link using constellations of $M=16$ and $M=64$, with probability dispersion parameters $\nu_{16}=0.215$ and $\nu_{64}=0.129$, respectively. These measurements span a range of modulation variances, allowing us to examine their effect on the resulting excess noise $\varepsilon$ and overall security of the protocol, assuming a reconciliation efficiency of $\beta = 95\%$.

Figure~\ref{fig:results-ALL} (\textbf{a}, \textbf{b}) presents the secret key rate (SKR) and excess noise as a function of modulation variance for the 16-state constellation. Each data point represents an average over multiple measurements. We compare the performance of the discrete modulation protocol, in terms of the average rate $\overline{\text{SKR}}$, with that of the CV Gaussian no-switching protocol~\cite{weedbrook2004quantum}, whose corresponding SKR is indicated by arrows in the plot. The average excess noise $\bar{\varepsilon}$ observed at the channel output is also reported, along with finite-size Gaussian bounds for parameter estimation with a misestimation probability $\varepsilon_{PE} = 10^{-10}$~\cite{ruppert2014long}.

Limiting the modulation variance $V_M$ to approximately 1 SNU or below ensures an acceptably low level of excess noise, as the effect of phase noise is minimal in this regime~\cite{hajomer2024long}. Notably, the performance gap between the 16-state discrete protocol and the continuous Gaussian modulation diminishes as $V_M$ is reduced to sub-SNU levels. At such low variances, using a 64-state constellation allows the discrete protocol to nearly match the performance of the continuous Gaussian protocol--see square markers in Fig.~\ref{fig:results-ALL} (\textbf{d}). The SKR difference between the latter two in this regime is less than $2\times10^{-4}$ bits/use, corresponding to under 1\% of relative key rate.

Figure~\ref{fig:results-ALL} (\textbf{c}–\textbf{e}) illustrates the interplay between system symbol rate, excess noise, and SKR for constellations with $M = 16$ (circles) and $M = 64$ (squares). As shown in Fig.~\ref{fig:results-ALL} (\textbf{c}), lower-order constellations consistently result in reduced excess noise, irrespective of the operating symbol rate. This trend is likely attributed to limitations in the DAC bits resolution, which more strongly affects higher-order modulations. Notably, operating at a low modulation variance maintains a relatively constant excess noise across different symbol rates, as evidenced by the overlapping error bars.

In terms of SKR in bit per channel use (Fig.\ref{fig:results-ALL} \textbf{d}), the 64-state constellation yields similar or higher key rates compared to the 16-state constellation, despite generally higher excess noise. However, when considering key throughput in bits per second (Fig.\ref{fig:results-ALL} \textbf{e}), the 16-state constellation operated at 16~GBaud demonstrates a significant advantage—achieving up to 60\% higher SKR. Here, SKRs are normalized to the highest-performing configuration reported (see Table\ref{tab:results-3}).

Table~\ref{tab:results-3} further underscores the practical impact of operating speed: even if higher-order modulations offer a SKR advantage in bit per channel use, system throughput is ultimately determined by symbol rate and implementation simplicity.

Accordingly, operating in the very low modulation variance regime presents a compelling trade-off: low-order constellations such as $M=16$ deliver comparable performance with the added benefits of robustness to implementation imperfections and high-speed operation. This highlights the practical value of discrete, low-order modulation formats for efficient and secure quantum key distribution under realistic system constraints.

Following this analysis, we focus on a 16-state constellation for measurements over a 20~km fiber link. We successfully extract a positive finite-size secure key rate, based on parameter estimation of the channel excess noise $\varepsilon$ and transmission loss $\eta$\cite{ruppert2014long}. A summary of the key parameters is provided in Table~\ref{tab:results-3}, and the theoretical loss limit for secure operation is plotted in Fig.\ref{fig:results-ALL} (\textit{f}), alongside results from the 10~km measurements for comparison.

\begin{table*}[]
\caption{\textbf{Experimental parameters and results}. $V_M$ modulation variance; $V_{el.}$: electronic noise; $ \varepsilon$: excess noise; $\beta = 95 \%$.}
\begin{tabular}{llllllllll}
\hline
\multicolumn{1}{|l|}{\begin{tabular}[c]{@{}l@{}}Cardinality \\ $M$ (a.u.)\end{tabular}} & \multicolumn{1}{l|}{\begin{tabular}[c]{@{}l@{}}Symbol rate \\ (Gbaud)\end{tabular}} & \multicolumn{1}{l|}{\begin{tabular}[c]{@{}l@{}}Link length\\ (km)\end{tabular}} & \multicolumn{1}{l|}{\begin{tabular}[c]{@{}l@{}}Channel loss\\ $\eta$ (a.u.)\end{tabular}} & \multicolumn{1}{l|}{\begin{tabular}[c]{@{}l@{}}$V_M$ \\ (SNU)\end{tabular}} & \multicolumn{1}{l|}{\begin{tabular}[c]{@{}l@{}}$V_{el.}$\\ (SNU)\end{tabular}} & \multicolumn{1}{l|}{\begin{tabular}[c]{@{}l@{}}$ \varepsilon$\\ (mSNU)\end{tabular}} & \multicolumn{1}{l|}{\begin{tabular}[c]{@{}l@{}}Number of \\ symbols\end{tabular}} & \multicolumn{1}{l|}{\begin{tabular}[c]{@{}l@{}}Asymptotic \\ SKR \\ (bits/use)\end{tabular}} & \multicolumn{1}{l|}{\begin{tabular}[c]{@{}l@{}}Finite-size \\ SKR \\ (Gb/s)\end{tabular}} \\ \hline
16  & 16  & 10  & 0.519 & 0.667  & 0.150  & 7.3 & $2\times 10^6$  & 0.0425  & 0.534 \\
64  & 10  & 10  & 0.514 & 0.703  & 0.112  & 8.0 & $2\times 10^6$  & 0.0486  & 0.400 \\
16  & 8   & 10  & 0.512 & 1.01   & 0.081  & 3.7 & $2\times 10^6$  & 0.0669  & 0.438 \\
16  & 16  & 20  & 0.346 & 0.513  & 0.151  & 5.7 & $2 \times 10^7$ & 0.0181  & 0.246        
\end{tabular}
\label{tab:results-3}
\end{table*}

\begin{table*}[t]
    \centering
    \caption{\textbf{A List of Integrated CVQKD Systems}}
    \label{tab:table1}
    \resizebox{0.8\textwidth}{!}{%
    \begin{tabular}{|c|c|c|c|c|}
        \hline
        \textbf{Ref.} & \textbf{Integrated Parts} & \textbf{Modulation} & \textbf{Platform} & \textbf{Symbol rate, GB} \\
        \hline
        \cite{li2023continuous} & Component (laser) & Gaussian & III-V/$\mathrm{Si_{3}N_{4}}$ & 0.25 \\
        \hline
        \cite{bian2024highly} & Component (VOA) & Gaussian & Si & 0.005 \\
        \hline
        \cite{Aldama:25} & Subsystem (Tx without laser) & Gaussian & InP & 0.016 \\
        \hline
        \cite{pietri2024experimental} & Subsystem (Balanced detector) & Gaussian & Si & 0.1 \\
        \hline
        \cite{bian2024continuous} & Subsystem (Balanced detector) & Gaussian & Si & 1 \\
        \hline
        \cite{hajomer2024continuous} & Subsystem (Phase-diverse receiver) & Discrete & Si/GaAs & 10 \\
        \hline
        \cite{zhang2019integrated} & Full system  without laser  & Gaussian & Si & 0.001 \\
        \hline
         Current work & Full system  without laser  & Discrete & Si & 16 \\
        \hline
    \end{tabular}%
    }
\end{table*}

\section{Discussion}
The large-scale deployment of secure quantum communication hinges on the chip-based QKD, which offers substantial advantages in cost reduction, miniaturization, and enhanced performance. DM-CVQKD stands out as a promising candidate for integrated photonic and electronic implementations, capable of achieving ultra-high symbol rates due to its compatibility with standard telecommunications technologies~\cite{hajomer2024continuous}. In this work, we have demonstrated the fastest integrated silicon photonic DM-CVQKD system, to date, operating at a symbol rate of 16 GBaud. This milestone was made possible through the integration of a broadband silicon photonic transmitter circuit coupled with integrated silicon photonic and electronic receiver circuits and advanced DSP.

Table \ref{tab:table1} summarizes previous attempts to integrate CVQKD systems. Most prior efforts have concentrated on integrating individual components or subsystems, with only one full silicon-based system demonstrated previously, which notably lacked an integrated laser source and was limited by a bandwidth of just 10 MHz, severely restricting its key rate to 250 kbit/s. In contrast, our integrated system provides both photonic and electronic integration and significantly exceeds previous demonstrations in terms of operational speed and overall system performance. Although our current implementation does not include an integrated VOA, modulation variance can alternatively be managed by controlling the driving voltage of the DAC, offering a practical solution within the system.

Compared to the previous records demonstration of high-rate DM-CVQKD \cite{wangSubGbpsKeyRate2022,hajomer2024continuous} our system significantly increases the symbol rate and doubles the achievable transmission distance. Furthermore, our system operates at room temperature, in contrast to high-speed discrete-variable (DV) QKD systems~\cite{li2023high, grunenfelder2023fast}.  Despite these advancements, several opportunities remain for further improvements. On the photonic integration front, integrating a laser source remains a key next step. Promising approaches for laser integration on silicon platforms, such as transfer printing technology~\cite{justice2012wafer}, could greatly enhance system compactness and functionality. To further advance compatibility with CMOS fabrication processes, future receiver designs will transition from GaAs pHEMT to Si bipolar or CMOS technology.

In terms of security analysis, two critical aspects warrant further exploration: accounting for non-Gaussian channels and extending existing analytical security proofs to achieve composable security. Additionally, achieving real-time operation is essential for practical deployment. A primary bottleneck currently lies in high-speed information reconciliation, which could potentially be addressed using GPU clusters.

To this end, our work represents a significant advancement toward the practical realization of low-cost, ultra-high-rate QKD systems, paving the way for the widespread deployment of secure quantum communications.

\begin{backmatter}

\bmsection{Acknowledgments} 
AAEH, ULA and TG acknowledge support from the Danish National Research Foundation, Center for Macroscopic Quantum States (bigQ, DNRF142). This project was funded within the QuantERA II Programme (project CVSTAR) that has received funding from the European Union’s Horizon 2020 research and innovation programme under Grant Agreement No 101017733. ID acknowledges support from the project 22-28254O of the Czech Science Foundation. We acknowledge support from  European Union’s Horizon Europe research and innovation programme under the project ``Quantum Security Networks Partnership'' (QSNP, grant agreement no. 101114043), from the European Union’s Digital Europe programme (QCI.DK, grant agreement no. 101091659 and BeQCI, grant agreement no. 101091625), and from Innovation Fund Denmark (CyberQ, grant agreement no. 3200-00035B).

\bmsection{Data availability} Data underlying the results presented in this paper are available from the authors upon reasonable request.

\smallskip

\bmsection{Disclosures} The authors declare no conflicts of interest.

\bmsection{Author contributions statement}

\bigskip

\end{backmatter}

% Bibliography
\bibliography{sample}

% Full bibliography added automatically for Optics Letters submissions; the following line will simply be ignored if submitting to other journals.
% Note that this extra page will not count against page length
\bibliographyfullrefs{sample}

%Manual citation list
%\begin{thebibliography}{1}
%\bibitem{Zhang:14}
%Y.~Zhang, S.~Qiao, L.~Sun, Q.~W. Shi, W.~Huang, %L.~Li, and Z.~Yang,
 % \enquote{Photoinduced active terahertz metamaterials with nanostructured
  %vanadium dioxide film deposited by sol-gel method,} Opt. Express \textbf{22},
  %11070--11078 (2014).
%\end{thebibliography}

% Please include bios and photos of all authors for aop articles
\ifthenelse{\equal{\journalref}{aop}}{%
\section*{Author Biographies}
\begingroup
\setlength\intextsep{0pt}
\begin{minipage}[t][6.3cm][t]{1.0\textwidth} % Adjust height [6.3cm] as required for separation of bio photos.
  \begin{wrapfigure}{L}{0.25\textwidth}
    \includegraphics[width=0.25\textwidth]{john_smith.eps}
  \end{wrapfigure}
  \noindent
  {\bfseries John Smith} received his BSc (Mathematics) in 2000 from The University of Maryland. His research interests include lasers and optics.
\end{minipage}
\begin{minipage}{1.0\textwidth}
  \begin{wrapfigure}{L}{0.25\textwidth}
    \includegraphics[width=0.25\textwidth]{alice_smith.eps}
  \end{wrapfigure}
  \noindent
  {\bfseries Alice Smith} also received her BSc (Mathematics) in 2000 from The University of Maryland. Her research interests also include lasers and optics.
\end{minipage}
\endgroup
}{}

\end{document}